\documentclass[12pt]{aa}
\usepackage[english]{babel}
\usepackage{graphicx}
\usepackage{longtable}

\onecolumn

\begin{document}

\title{Instability of the kinematic state in the atmosphere of the hypergiant $\rho$\,Cas outside outburst}

\author{V.G.~Klochkova$^1$, V.E.~Panchuk$^1$, N.S.~Tavolganskaya$^1$, and I.A.~Usenko$^2$}

\date{\today}

\institute{1 -- Special Astrophysical Observatory,  Nizhnij Arkhyz, 369167, Russia \newline 
           2 --  Astronomical Observatory, Odessa National University, Odessa, 65014 Ukraine}

\abstract{Observations of the yellow hypergiant $\rho$\,Cas obtained in 2007--2011 in a wide
wavelength region with spectral resolution R\,$\ge$60\,000 have enabled studies of features of its
optical spectrum in detail and brought to light previously unknown characteristics of the extended
atmosphere of the star. The radial velocity measured from symmetric absorptions of metals varies with an
amplitude of about $\pm 7$\,km/s around the systemic velocity Vsys\,=\,$-47$\,km/s, due to low-amplitude pulsations
of the atmospheric layers near the photosphere. At some times, a velocity gradient was found in deep
atmospheric layers of the star. A slight velocity stratification in the stellar atmosphere was detected for the
first time, manifested as a difference of 3--4 km/s in the velocities measured from absorption lines of neutral
metals and of ions. The long-wavelength components of split absorptions of BaII, SrII, TiII, and other
strong lines with low excitation potentials for their lower levels are distorted by nearby emission lines. 
It is suggested that the short-wavelength components, whose locations correspond to the narrow velocity 
range Vr(blue) from approximately $-60$ to $-70$\,km/s, are formed in a circumstellar envelope; one component of
the D~NaI doublet and the emission components of the FeII\,6369.46 and 6432.68\,\AA{} ions are also formed 
there.\newline
}

\titlerunning{\it  Kinematic state in the atmosphere of  $\rho$ Cas}
\authorrunning{\it Klochkova et al.}
\maketitle

\section{Introduction}

Yellow hypergiants  -- rare stars whose prototype is $\rho$\,Cas (Sp\,=\,G2\,Iae) are evolved high-mass stars
with super-high luminosities. Such objects are found near the luminosity limit on the Hertzsprung--Russell
diagram, in the instability region containing hypergiants with spectral types from A to M [\cite{Jager1998,Jager2001,Hump}]. 
The abundances of CNO elements and the sodium excess in the atmosphere of $\rho$\,Cas [\cite{Takeda}] indicate that 
the star has already passed the red-supergiant phase, and is now in a rapid evolutionary transition from a red
supergiant to a Wolf--Rayet or LBV star. The process of this motion towards higher temperatures is poorly
studied.

Apart from their high luminosities, yellow hypergiants differ from ordinary supergiants in their high
mass-loss rates via their stellar winds and in the presence of circumstellar envelopes. The instability of
these objects is also manifest through pulsation-type spectral and brightness variations. The characteristic
features of the pulsations of massive stars at the stage of contraction of the helium core were considered by
Fadeyev [\cite{Fadeev}], who concluded that long-period radial pulsations were improbable for $\rho$\,Cas. Along with the
above manifestations of instability, yellow hypergiants also undergo so-called ``shell episodes'', when the
star loses matter especially efficiently and becomes enshrouded for several hundred days by the ejected cool matter, 
which forms a pseudo-photosphere. In the case of $\rho$\,Cas, the most recent event of this kind
occurred in late 2000 -- early 2001, when the star lost up to $3\times 10^{-2} \mathcal{M}_{\sun}$ [\cite{Lobel}].

On the Hertzsprung--Russell diagram, $\rho$\,Cas is at the boundary of the so-called ``yellow void'' [\cite{Jager1998}]
separating hypergiants from LBV stars in quiescence. Hypergiants near the void exhibit negative density
gradients and nearly-zero surface gravities log\,g [\cite{Nieuw}], favoring the formation of an instability region in the
atmosphere. The pulsation amplitudes of yellow hypergiants apparently strongly increases at the boundary of the yellow 
void, resulting in increased atmospheric instability and shell ejection [\cite{Jager1998}]. 

Close to $\rho$\,Cas in the Hertzsprung--Russell diagram, we find the hypergiant V1302\,Aql, better
known as the associated IR--source IRC\,+10420. The central star of V1302\,Aql (spectral type F8\,Ia, 
luminosity of the order of $10^6 L_{\sun}$) is surrounded by a dense gas and dust medium, so that only the wind can be
observed. Despite its fairly high effective temperature, the star is associated with a strong OH maser. One
of the most important results from many-year studies of V1302\,Aql is the discovery of a rapid growth of its
effective temperature [\cite{IRC10420,Oudm}]. Monitoring data from the last decades of the 20\,th century indicate 
an acceleration of this temperature increase [\cite{IRC}]. In contrast to the hypergiant V1302\,Aql, which has a massive
and structured envelope, $\rho$\,Cas is point-like in observations. An envelope far from the star that would
be detectable with the Hubble Space Telescope is not present in  $\rho$\,Cas, providing evidence that the duration
of the high mass-loss stage has been short [\cite{Schuster}]. The peculiarity of  $\rho$\,Cas and its spectral variations
were detected more than a century ago (cf. [\cite{Bidelman}] and references therein); however we still do not have a
complete understanding of the physical processes resulting in the complex, time-variable kinematic situation in 
the extended atmosphere of this hypergiant. Fadeyev [\cite{Fadeev}] notes that even the pulsation type remains
unclear, making monitoring important.

Our two decade of spectroscopic monitoring of the hypergiant V1302\,Aql with the 6-m telescope of the
Special Astrophysical Observatory  enabled us to draw conclusions concerning its evolution [\cite{IRC10420,IRC}]. 
The similarity of the evolutionary stage and fundamental parameters of V1302\,Aql to those of $\rho$\,Cas stimulated 
monitoring for the latter rare yellow hypergiant. The current paper presents the results of
our optical spectroscopic observations of $\rho$\,Cas performed in 2007--2011. Section~2 briefly describes the
observations and data analysis. Section~3 presents our results and compares them to those published
earlier. Our main conclusions are given in Section~4.

\section{Observations, reduction, and spectral analysis}

Spectroscopic observations of $\rho$\,Cas were carried out at the Nasmyth focus of the 6-m telescope using 
the NES echelle spectrograph [\cite{NES1,NES2}]. The mean epochs of our observations (JD) and the
recorded spectral ranges are listed in Table\,\ref{obs}. The observations were performed using a 
2048$\times$2048\,pixel CCD chip and an image slicer [\cite{NES2}].  A 2K$\times$4K
CCD chip was used on September~14,~2011. Table\,\ref{obs} shows that this transition to a large-format
CCD considerably expanded the recorded wavelength range, $\Delta \lambda$. The spectroscopic resolving power
was $\lambda/\Delta \lambda \ge 60000$, with the signal-to-noise ratio $S/N\ge$100.

The one-dimensional spectra were extracted from the two-dimensional echelle frames using the modified [\cite{Yushkin}] 
ECHELLE routine of the MIDAS software package. To remove cosmic-ray traces, we applied median averaging of two spectra taken just after
one another. Our wavelength calibration was done using spectra from a Th\,Ar hollow-cathode lamp.
We checked the instrumental agreement between the stellar and hollow-cathode lamp spectra using
O$_2$ and H$_2$O telluric lines. To check the derived Vr values, we measured 15--20 telluric lines in the
spectra of  $\rho$\,Cas with their long-wavelength end at 5930--6010\,\AA{} and up to 70--80 lines in spectra 
with longer-wavelength ends. The rms uncertainty in the Vr measurements of narrow telluric absorption lines
is $\le0.5$\,km/s (the uncertainty for a single line). The accuracy is somewhat poorer in the case of $\rho$\,Cas,
since the spectral lines are broadened by turbulence: the microturbulence velocity in the atmosphere
reaches 11\,km/s [\cite{Jager1998}]. The techniques used for the Vr measurements derived from spectra taken with
the NES spectrograph, their uncertainties, and the sources of these uncertainties are described in more
detail in [\cite{RV1,RV2}].

\section{Main results}

\section{Effective temperature of $\rho$\,Cas}

We determined the effective temperature, Teff, using the spectroscopic criteria developed by Kovtyukh [\cite{Kovt}]. 
This method is based on ratios of selected spectral lines that are sensitive temperature
indicators. A single pair of lines provides temperature with an uncertainty of 50--110\,K, but using a set of
criteria results in a fairly accurate mean value. We are able to use more than 100 line pairs in spectra of
F--G stars, making it possible to reach an internal accuracy of 10--30\,K. Table\,\ref{obs} shows that, due
to the large line widths in the case of $\rho$\,Cas, our uncertainty is 39--53\,K for the spectra in the range
5200--6700\,\AA{}.
Due to a smaller number of available line pairs  at $\approx$4400--6020\,\AA{}  the accuracy is poorer
in this range: 90--170\,K. Table\,\ref{obs} shows that the effective temperature varied during our observations
in the range 5777--6744\,K, with the mean being about 6200\,K. The temperature variations we detected for   
$\rho$\,Cas in the course of its pulsation period exceed the temperature difference $\Delta$Teff$\approx$750\,K 
obtained earlier by Lobel et al. [\cite{Lobel1998}].

\section*{Peculiarity and profile variations of spectral features}

\begin{table}
\bigskip
\caption{Log of observations of $\rho$\,Cas and the effective temperatures Teff derived for various dates}
\begin{tabular}{ c| c| c| c| l}  
\hline
Spectrum &Date &JD & $\Delta\lambda,$ &\hspace{0.2cm} \small Teff,   \\
No.      &  &2450000+ & \AA{} &\hspace{0.3cm} \small K      \\
\hline     
s493015& 09.03.2007&4168.63  & 4557--6014 &6221$\pm$90  \\
s494023& 10.03.2007&4169.57  & 4557--6014 &6200$\pm$171 \\ 
s495019& 10.03.2007&4170.49  & 4514--5940 &6229$\pm$131 \\
s516015& 21.02.2008&4518.39  & 5204--6680 &6610$\pm$53  \\
s525032& 19.10.2008&4759.23  & 3050--4520 &             \\
s526006& 20.10.2008&4760.23  & 5214--6690 &6744$\pm$53  \\
s538009& 30.09.2009&5104.62  & 5216--6691 &6420$\pm$39  \\
s553018& 01.08.2010&5409.52  & 4422--5930 &5777$\pm$161 \\
s554032& 23.09.2010&5463.39  & 3970--5390 &             \\
s555027& 24.09.2010&5464.39  & 5216--6690 &6044$\pm$40  \\
s564020& 13.01.2011&5574.60  & 5208--6683 &6174$\pm$43  \\
s565003& 13.01.2011&5575.09  & 5208--6683 &6322$\pm$52  \\
s575002& 14.09.2011&5819.41  & 3985--6980 &             \\
\hline
\end{tabular}
\label{obs}
\end{table}

\begin{figure}
 \includegraphics[angle=0,width=0.5\textwidth,bb=40 60 570 790,clip]{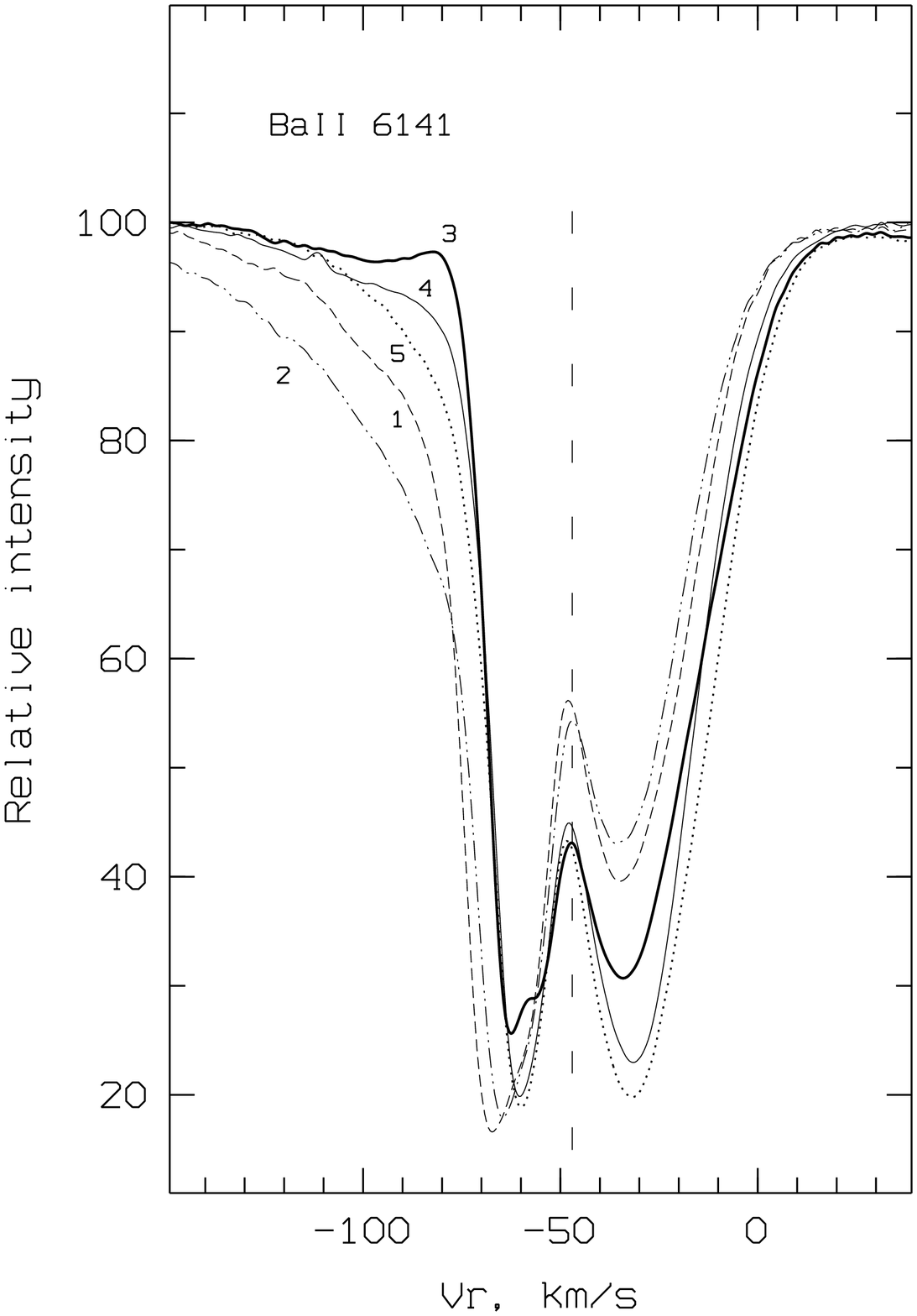} 
 \includegraphics[angle=0,width=0.5\textwidth,bb=40 60 570 790,clip]{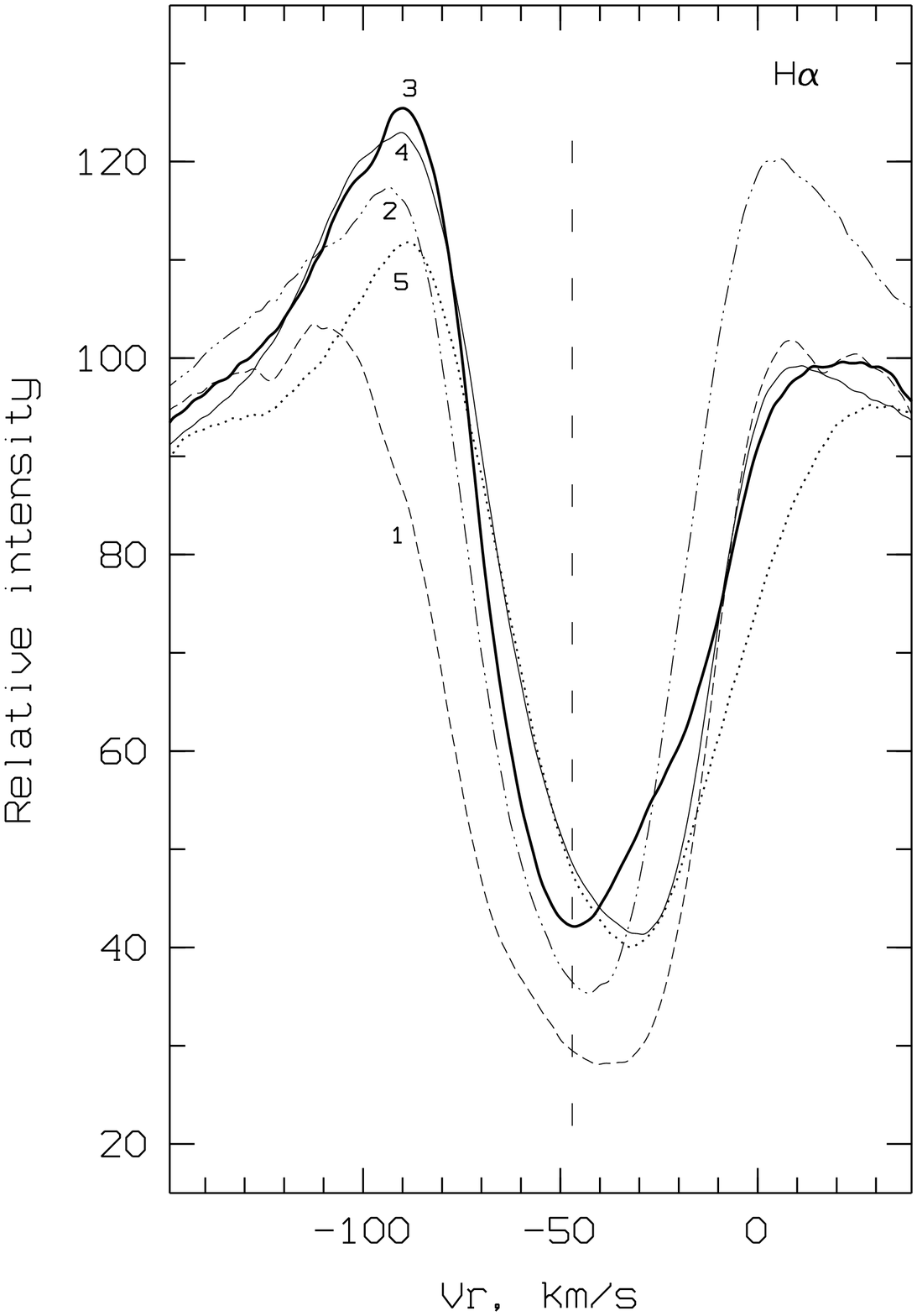}
 \caption{BaII\,6141\,\AA{}  profile and the central part of the  H$\alpha$ profile 
      in the spectra of $\rho$\,Cas for different epochs: 1 --  February~21,~2008, 2 -- October~20,~2008, 
      3 -- September~30,~2009, 4 -- September~24,~2010, 5 -- January~13,~2011. 
      The vertical dashed line indicates the systemic velocity, Vsys\,=\,$-47$\,km/s [\cite{Lambert,Lobel1998}].}
 \label{fig1}  
 \end{figure}

The profiles of strong absorption lines in the spectrum of $\rho$\,Cas are variable and, as a rule, asymmetric:
their short-wavelength wings are either raised above the continuum by variable emission or are more extended 
compared to the long-wavelength wing. A good example is the profile of the BaII\,6141\,\AA{} line
presented in the upper panel of Fig\,\ref{fig1}, which plots the residual intensity vs. the heliocentric radial velocity
Vr. The short-wavelength wing of the BaII\,6141\,\AA{} line, which forms in upper atmospheric layers under
the influence of the stellar wind, sometimes attains values near $-120$\,km/s, even reaching $-170$\,km/s on
October~20,~2008. Variations in the extended short-wavelength wings reflect kinematic instability in the
upper levels of the atmosphere, which are subject to influence of the stellar wind.

In addition to extended short-wavelength wings, the strongest absorption lines in the spectra of $\rho$\,Cas
also feature a peculiarity in their cores, which are permanently split into two components. The splitting of strong 
low-excitation absorption lines in the spectrum of $\rho$\,Cas has been known for a long time
(cf. references in [\cite{Lambert}]). More than half a century ago, Sargent [\cite{Sargent}] presented a long list of 
such spectroscopic features. The wide spectral range recorded by us enabled us to identify 10--12 split absorption
lines in the visible range, and more than 100 in the extended wavelength range, at $\lambda > 3500$\,\AA{}. A list 
of split absorption lines with the most trustworthy positions  at $\lambda > 3900$\,\AA{} is presented in 
Table\,2.  Numerous split features are also present in the shorter wavelength range, $\lambda$\,=\,3500--3900\,\AA{},
but their measurements are hindered by the fact that the spectrum is rich with blends. 
The lines with wavelengths longer than 5100\,\AA{}  were already listed by Sargent [\cite{Sargent}];
the shorter-wavelength split absorption lines in the spectrum of $\rho$\,Cas were identified by us. We consider
possible origin of this splitting in the next section, which deals with our analysis of the velocity field.

Our recorded spectral range contains the H$\alpha$ line at six epochs. The complex absorption and emission
H$\alpha$ profile varies in time: the lower panel of Fig.\,\ref{fig1} demonstrates that the position of the absorption core,
intensities of the emission components, and their intensity ratios are all time variable. The varying
position of the H$\alpha$ absorption core indicates that its formation region moves in the stellar atmosphere.
Note that the displacement of the H$\alpha$ core during our observations was $\approx$16\,km/s, while the earlier
observations by Lobel et al. [\cite{Lobel1998}] detected core displacements as large as 35\,km/s. No strict correlation
is observed in the evolution of the H$\alpha$ and BaII\,6141\,\AA{} profiles.

It was proposed in [\cite{Jager1977}] that the H$\alpha$ emission was formed in outer atmospheric layers that are thermally
excited by shock waves. The emission is formed in thin high-temperature layers behind the shock front.
Recombination occurs at this location, and the resulting emission lines should be seen against the
absorption profile when the amount of hydrogen ionization due to the passage of the shock becomes high
enough. The amount of ionization is related to the velocity of the shock. Note that the shock velocities
that must be provided for Balmer emission to appear are much higher than the velocities observed for
$\rho$\,Cas. In a model with a single shock, there is a slight velocity gradient in the recombination region, which
determines the width of the emission line (taking into account integration across the visible hemisphere).
It is emphasized in [\cite{Jager1977}] that the cases when the emission line width is determined by thermal motions
in a single high-temperature region and by several recombination regions formed at the boundary of each
turbulent element cannot be distinguished spectroscopically. For this reason, the assumed turbulence
spectrum is of key importance for explaining the observed microturbulence velocity, 11\,km/s.

Another effect whose explanation requires the assumption of an ensemble of shocks with recombination is 
the long-lived, virtually permanent, presence of emission features. For example, a high optical
depth of the heated layer in the L$\alpha$ line was used in [\cite{Gorb}] to explain the long-lived presence of Balmer
emission in Mira stars. It is noted in [\cite{Bychkov}] that it is difficult to explain long-lived Balmer emission in
models with a shock emerging in the atmosphere, since the high-temperature region is quickly cooled
by free--free transitions. Thus, the assumption of multiple shocks [\cite{Jager1977}] is also required to explain the
long duration of this emission. However, if indeed the observed line splitting is due to an emission component 
that ``pushes apart'' the remnant of a broad absorption line, we obtain an additional question: what
is the excitation mechanism for the selected bound--bound transitions?  We will consider the multipleshock 
hypothesis further in a special study. 

We noted in the Introduction  the similarity of the fundamental parameters and evolutionary stages for the two cool hypergiants  
$\rho$\,Cas and V1302\,Aql. However, the two stars also have important differences.
The first is the mass-loss rate: for V1302\,Aql, this is $3\div6\times 10^{-4} \mathcal{M}_{\sun}/$year, 
with possible episodes of increase to $10^{-3} \mathcal{M}_{\sun}/$year (see [\cite{Hump2002}] and references
therein). This is one to two orders of magnitude higher than the upper limit for the mass-loss rate of
$\rho$\,Cas, $9.2 \times 10^{-5} \mathcal{M}_{\sun}/$year [\cite{Lobel1998}]. The high mass-loss
rate of V1302\,Aql has resulted in the formation of the powerful, structured circumstellar envelope observed
by the Hubble Space Telescope [\cite{Hump2002}]. The presence of this envelope produces a large IR--excess, making
V1302\,Aql a bright IR--source. The IR--fluxes of this star in the IRAS bands are higher than those of
$\rho$\,Cas by a factor of 40--50. Despite the similarity of their MK spectral types, the optical spectra of
the two hypergiants are also considerably different. Permitted and forbidden emission lines, often with
intensities exceeding those of the local continuum by large factors, dominate the spectrum of V1302\,Aql.
The HI and CaII line profiles are two-peaked, and lines of metal ions often possess P\,Cygni profiles 
[\cite{IRC10420,Oudm,IRC}]. The spectrum of  $\rho$\,Cas resembles that observed for V1302 Aql in the 1970s 
[\cite{Hump1973}]: it is close to that of a normal F--supergiant, with a modest addition of
some emission features. The main differences from the spectra of normal supergiants are its broader 
absorption lines, due to very high luminosity and well-developed atmospheric turbulence, and the
splitting of the strongest metal absorption lines.

\section*{Velocity field in the atmosphere and envelope of  $\rho$\,Cas}

\section*{The systemic velocity of $\rho$\,Cas} 

Lambert et al. [\cite{Lambert}] used observations of the IR spectrum of $\rho$\,Cas to determine the heliocentric 
systemic velocity, Vsys\,=\,$-48\pm2$\,km/s. Lobel et al. [\cite{Lobel1994}] used the systemic velocity 
Vsys\,=\,$-42$\,km/s, while a later paper of Lobel et al. [\cite{Lobel1998}] assumes Vsys\,=$-47\pm 1$\,km/s.
We have adopted  Vsys\,=$-47\pm 1$\,km/s, in agreement with the results of [\cite{Lobel1998}]. 
Note that the membership of  $\rho$\,Cas in the stellar association Cas~OB5 suggests the mean velocity of a sample 
of 21 stars in the association, Vass\,=\,$-44.5$\,km/s [\cite{Bartaya}] for the systemic
velocity of $\rho$\,Cas.

 \section*{The radial velocity from symmetric absorption lines.} 

Lobel et al. [\cite{Lobel}] monitored the radial  velocity of  $\rho$\,Cas   using narrow fragments,  
45\,\AA{} wide, centered at 5187\,\AA{}. We analyzed the velocity field in the atmosphere of $\rho$\,Cas 
using large numbers of individual spectral features, a fundamental difference
from the approach used in [\cite{Lobel}]. Our wide wavelength range and the high accuracy of our 
velocity measurements from individual lines enabled us to study the velocity field using an 
unprecedentedly large number of isolated single lines (several hundred in each of the
spectra), as well as split lines.

 \section*{Components of split absorption lines.}

The second group of lines are the short-wavelength  components of split low-excitation absorption lines.
A list of split absorption lines in the spectra of $\rho$\,Cas is presented in Table\,2. 
The identification and wavelengths follow the data of the spectroscopic atlas [\cite{atlas}]; lower-level 
excitation potentials were taken from the VALD database [\cite{VALD1,VALD2}].

\begin{figure}[h!]
\includegraphics[angle=0,width=0.25\textwidth,bb=50 35 550 780,clip]{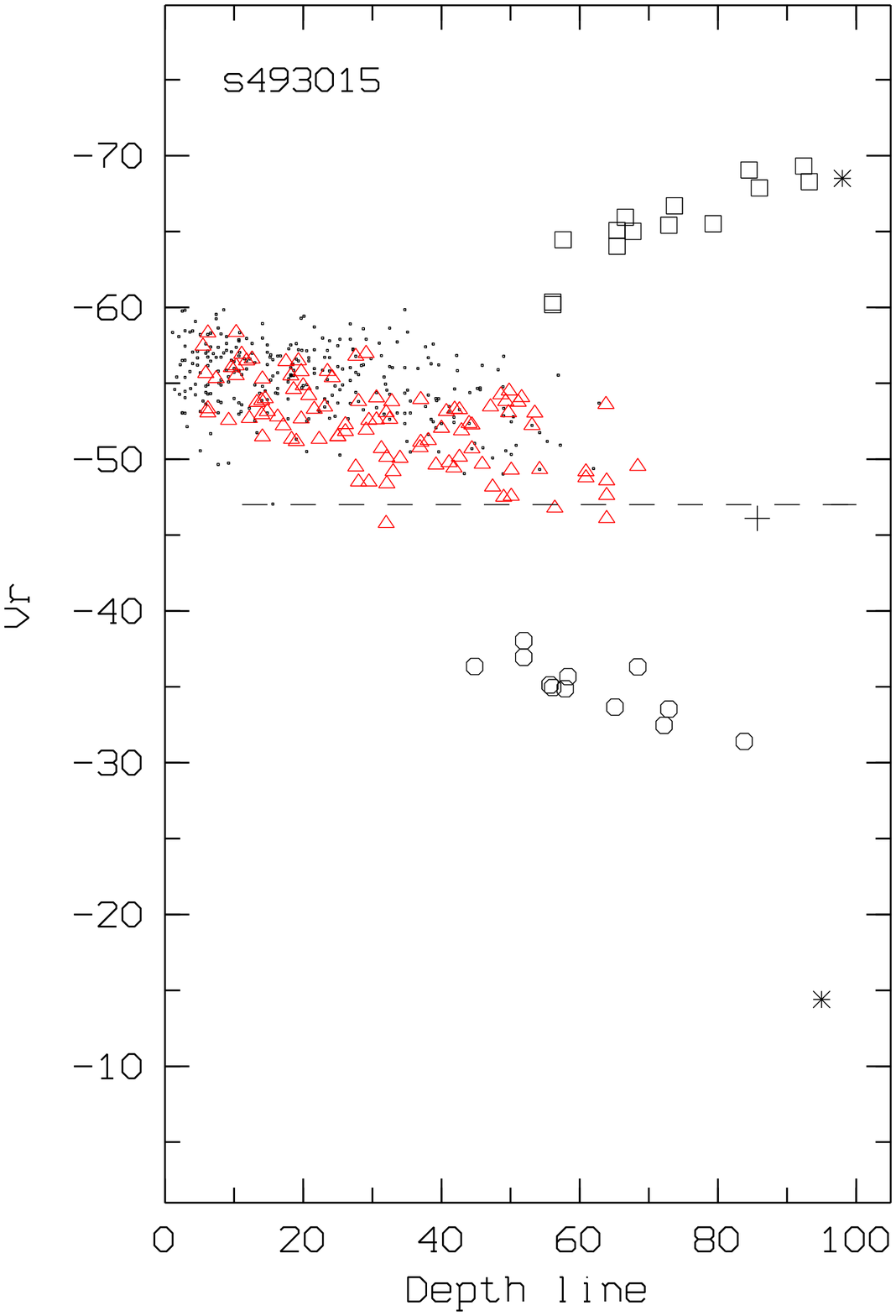} 
\includegraphics[angle=0,width=0.24\textwidth,bb=70 40 550 780,clip]{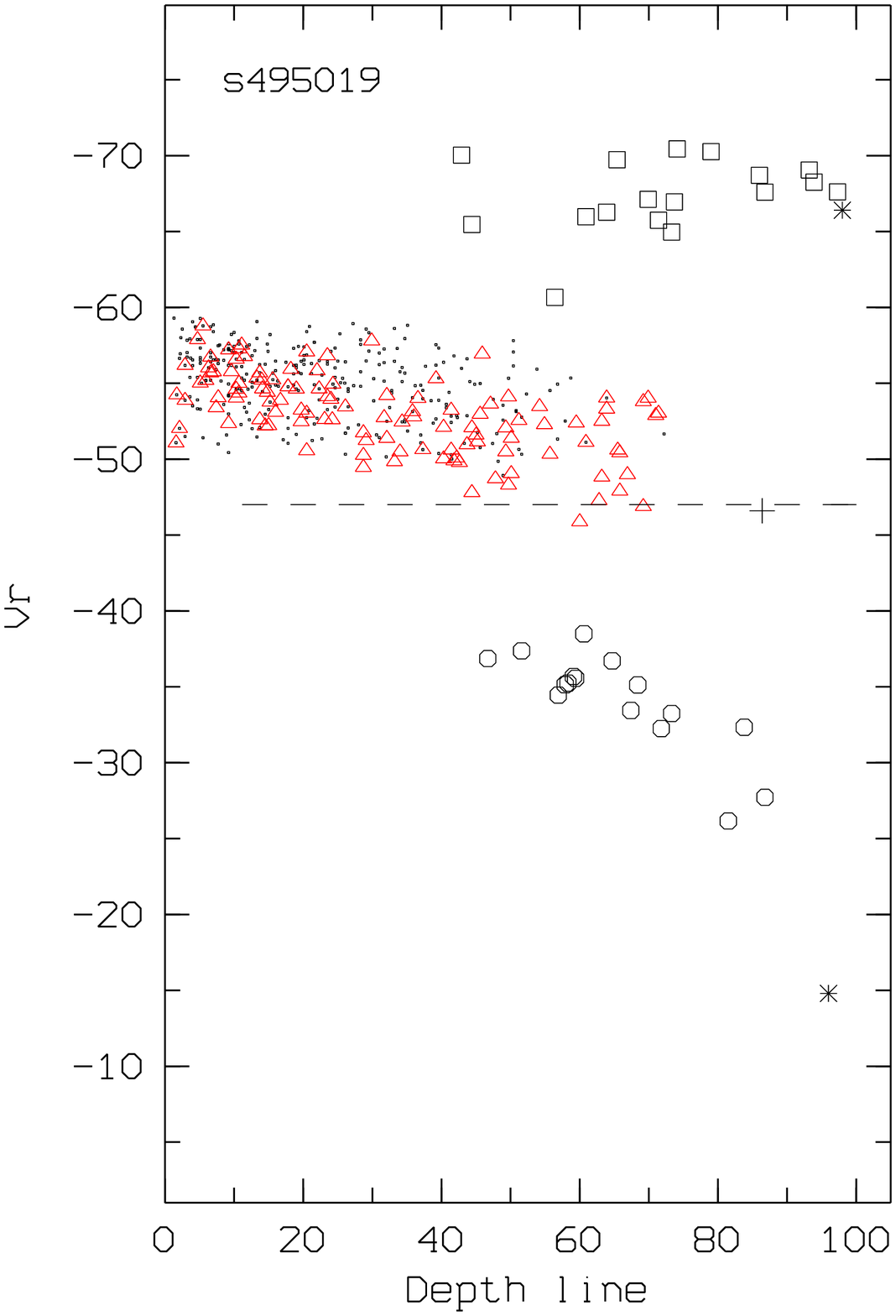} 
\includegraphics[angle=0,width=0.24\textwidth,bb=70 40 550 780,clip]{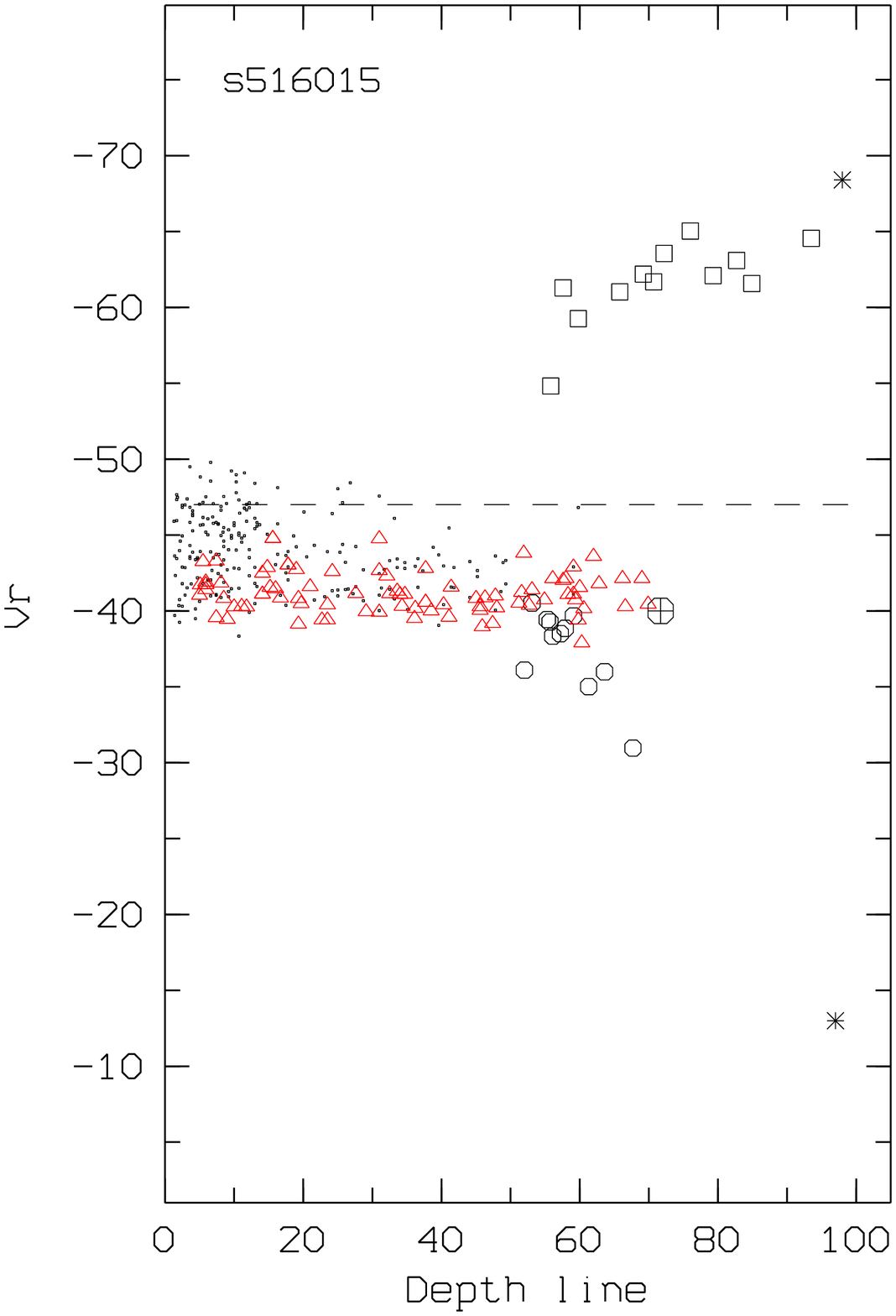} 
\includegraphics[angle=0,width=0.24\textwidth,bb=70 40 550 780,clip]{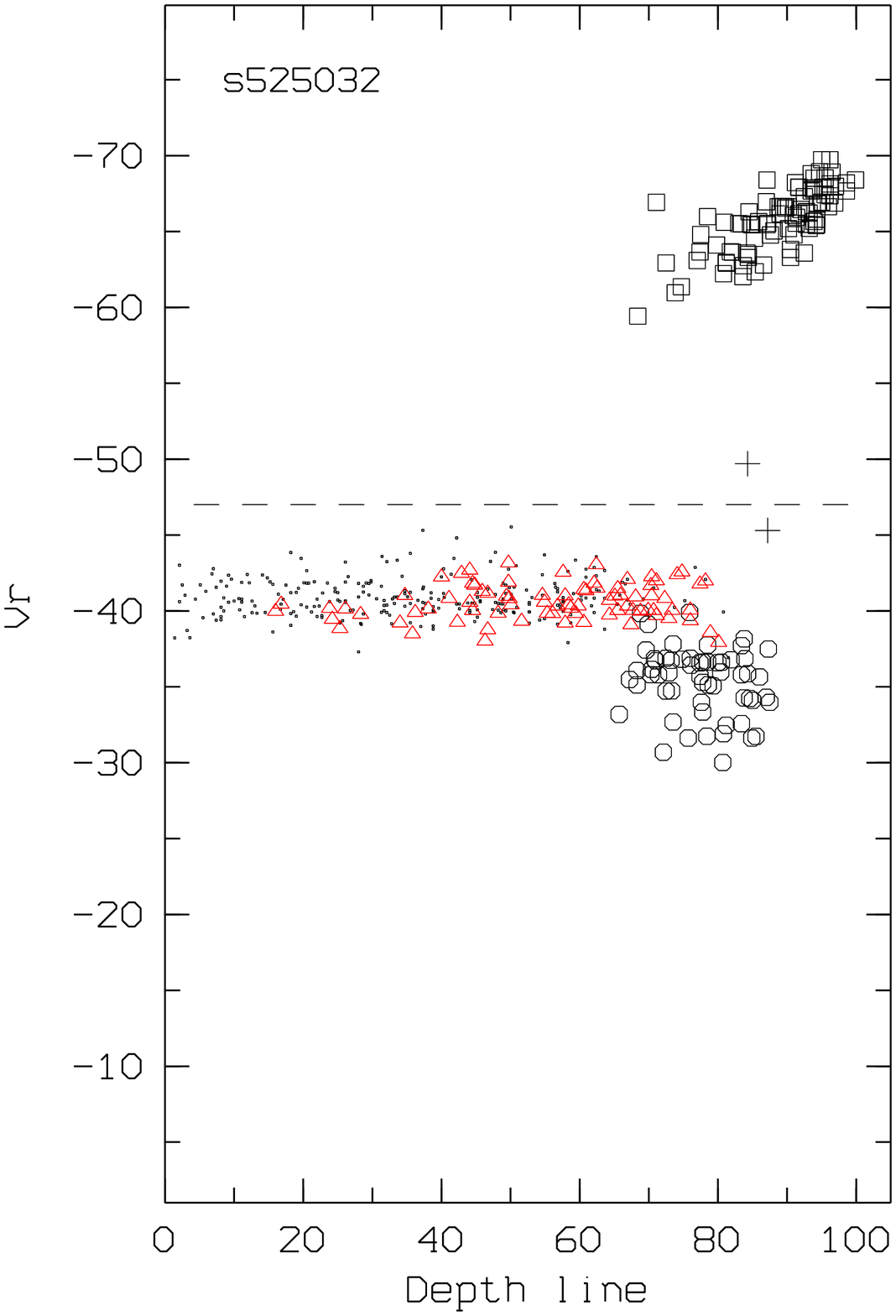} 
\includegraphics[angle=0,width=0.25\textwidth,bb=50 35 550 780,clip]{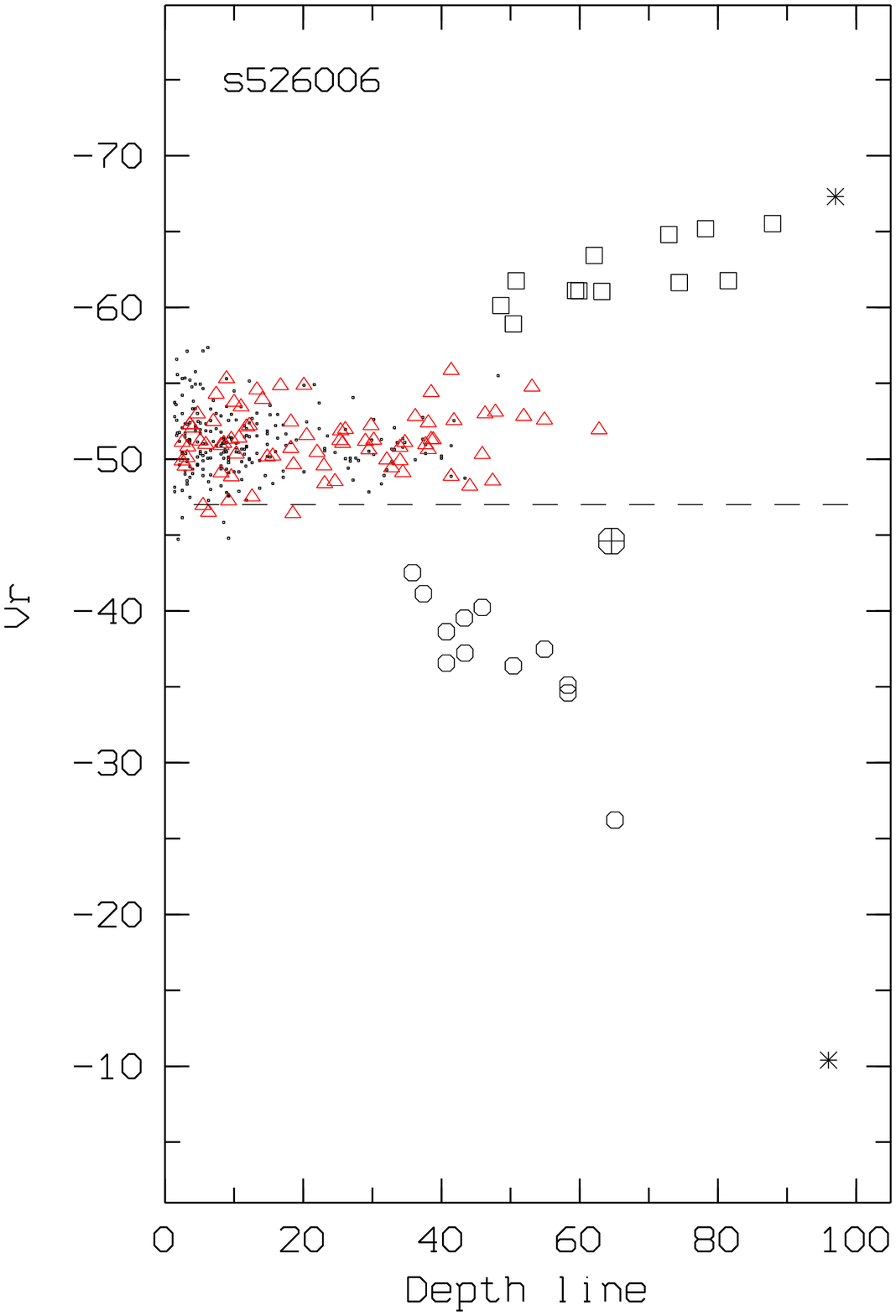} 
\includegraphics[angle=0,width=0.24\textwidth,bb=70 40 550 780,clip]{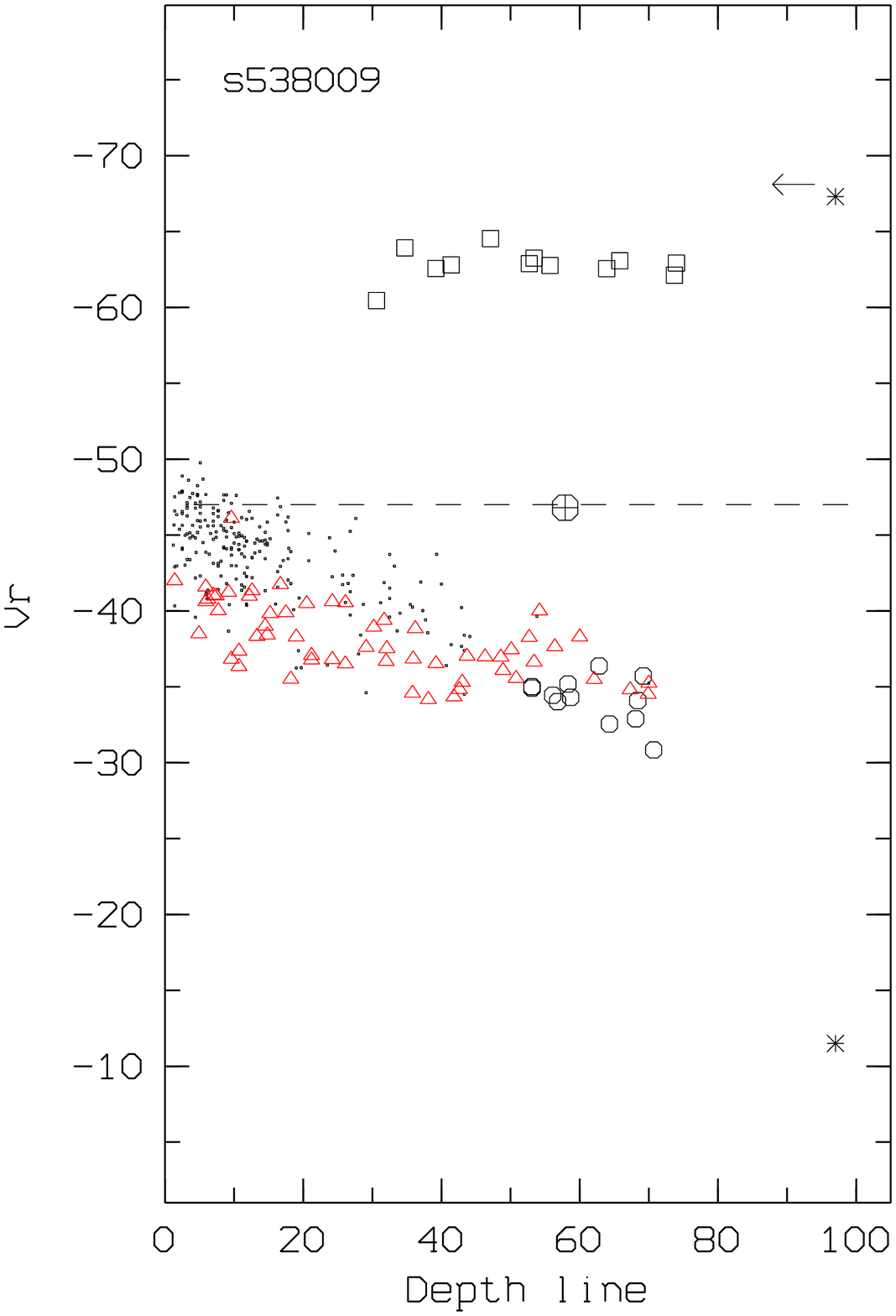} 
\includegraphics[angle=0,width=0.24\textwidth,bb=70 40 550 780,clip]{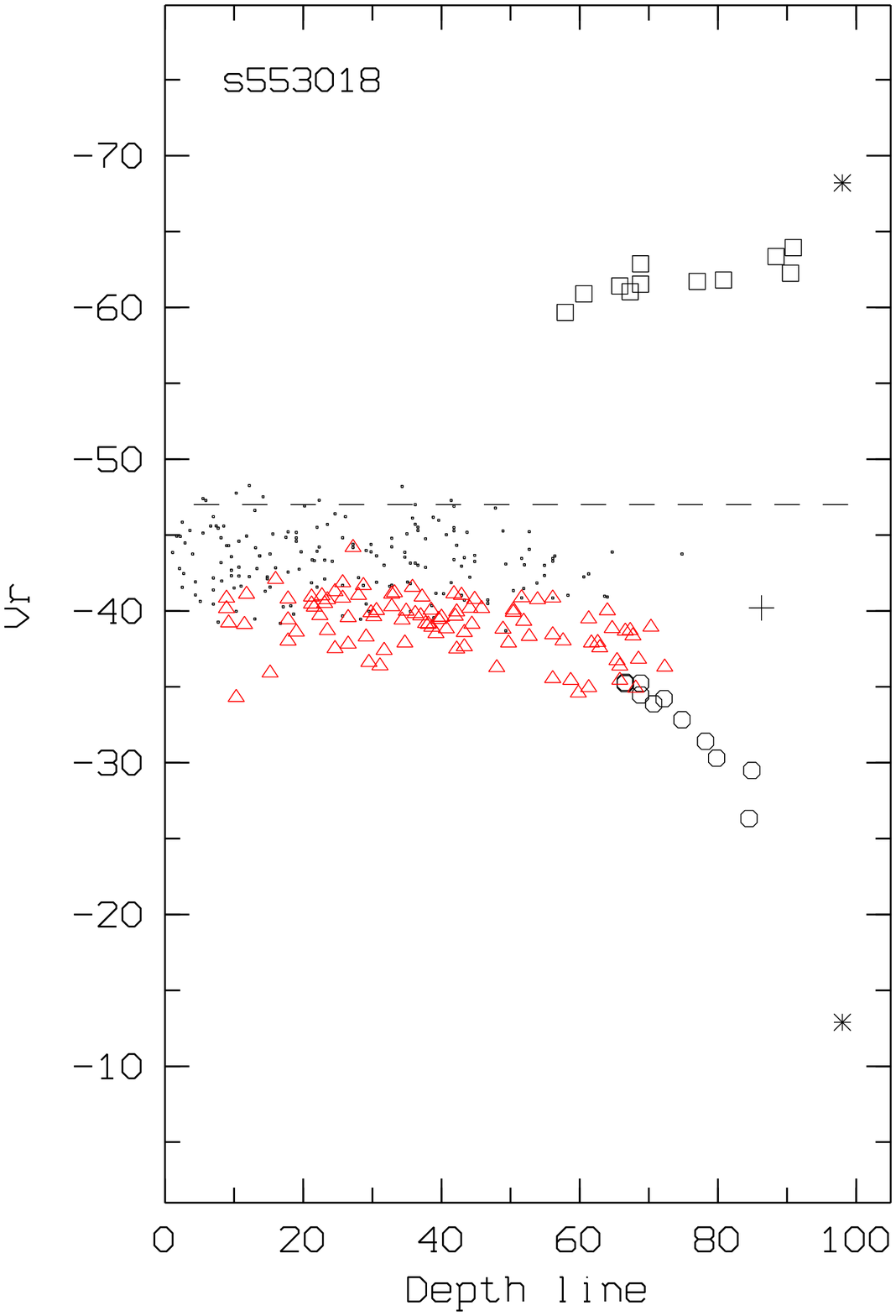} 
\includegraphics[angle=0,width=0.24\textwidth,bb=70 40 550 780,clip]{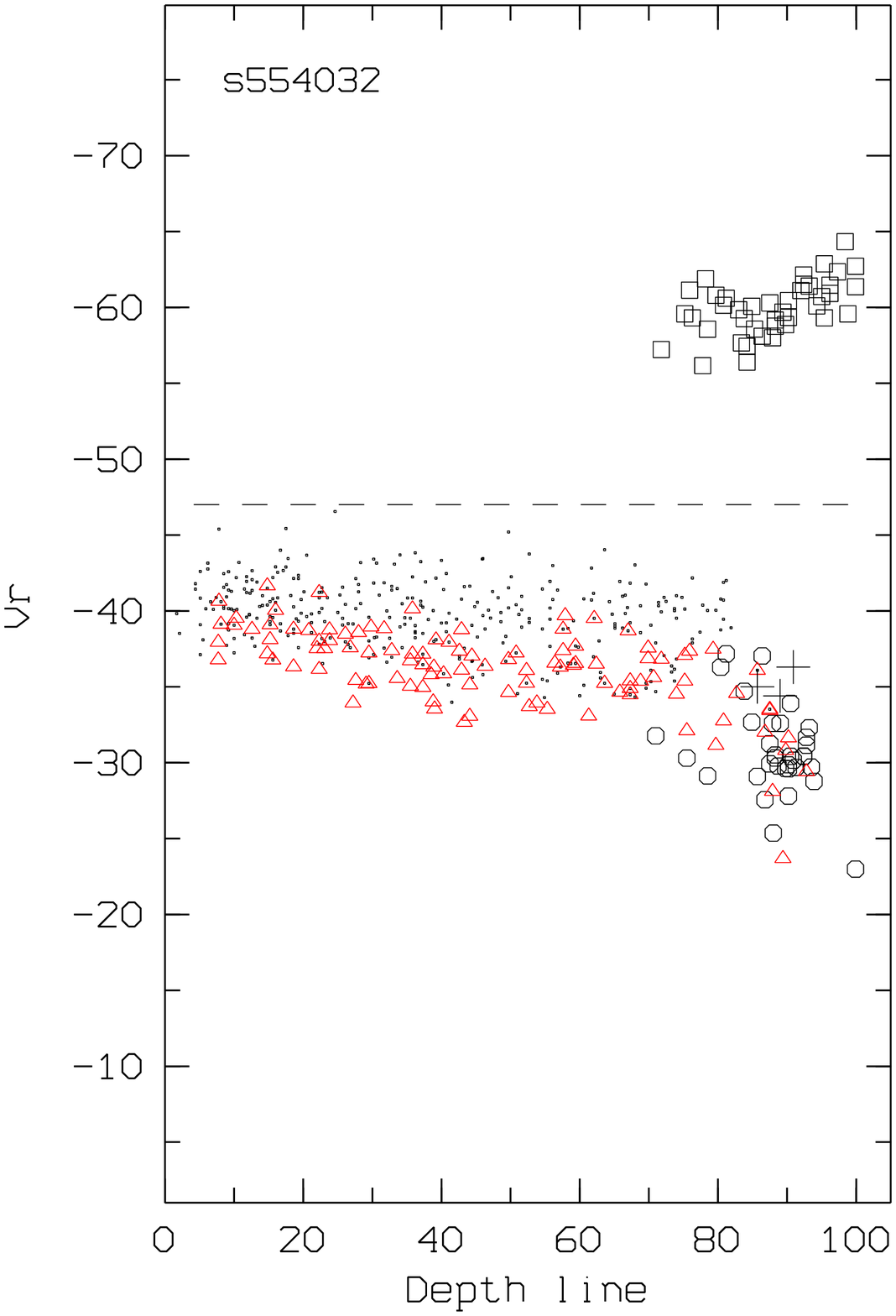} 
\includegraphics[angle=0,width=0.25\textwidth,bb=50 35 550 780,clip]{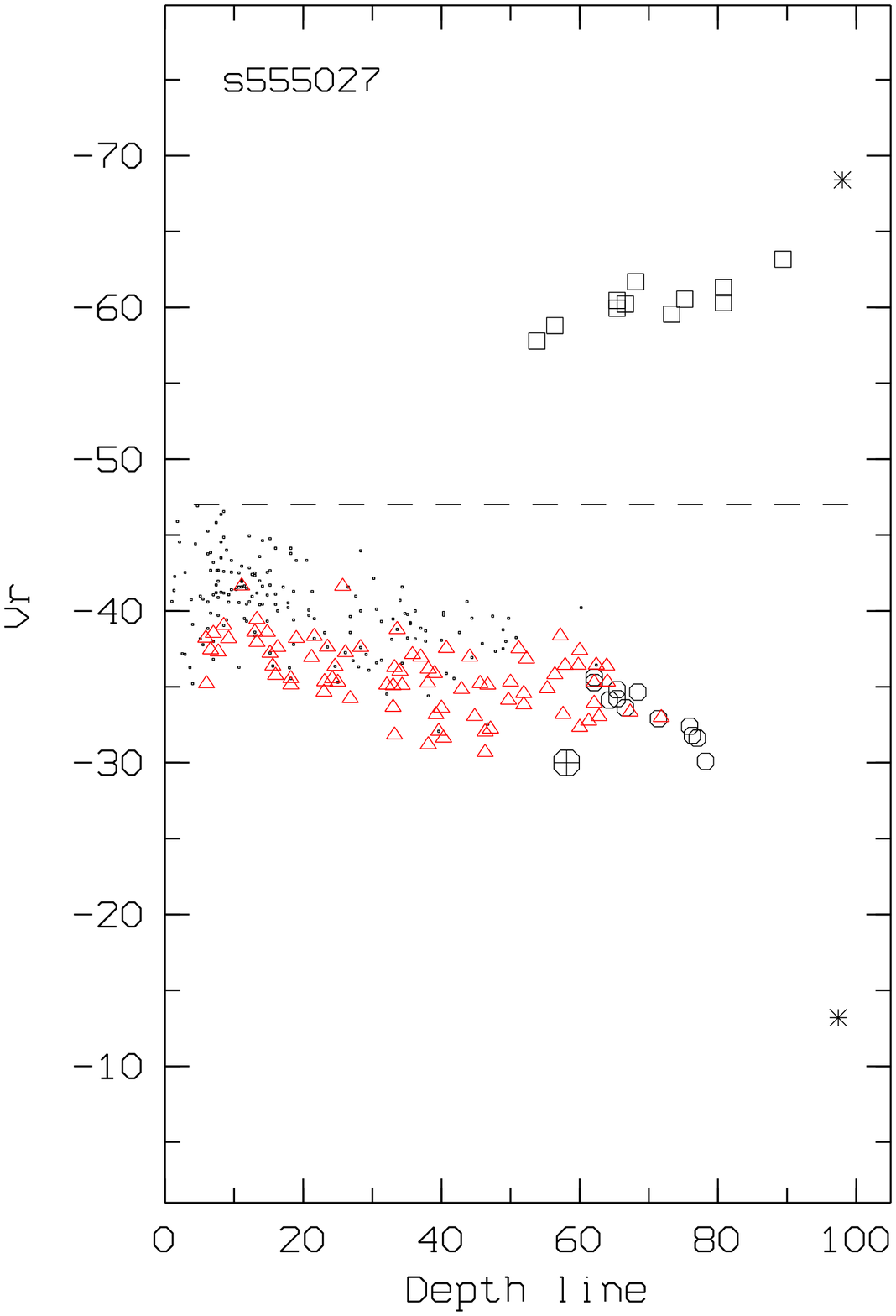} 
\includegraphics[angle=0,width=0.24\textwidth,bb=70 40 550 780,clip]{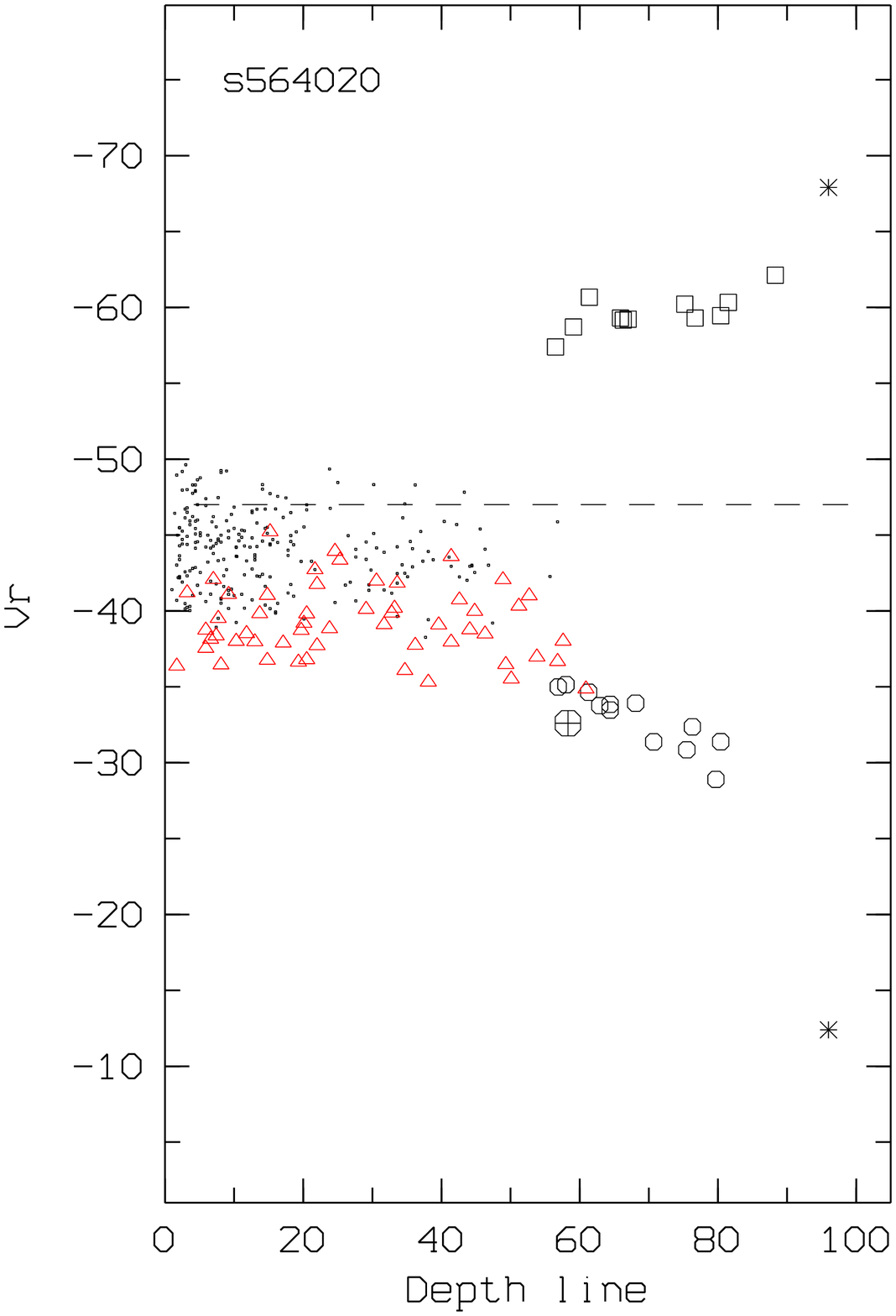} 
\includegraphics[angle=0,width=0.24\textwidth,bb=70 40 550 780,clip]{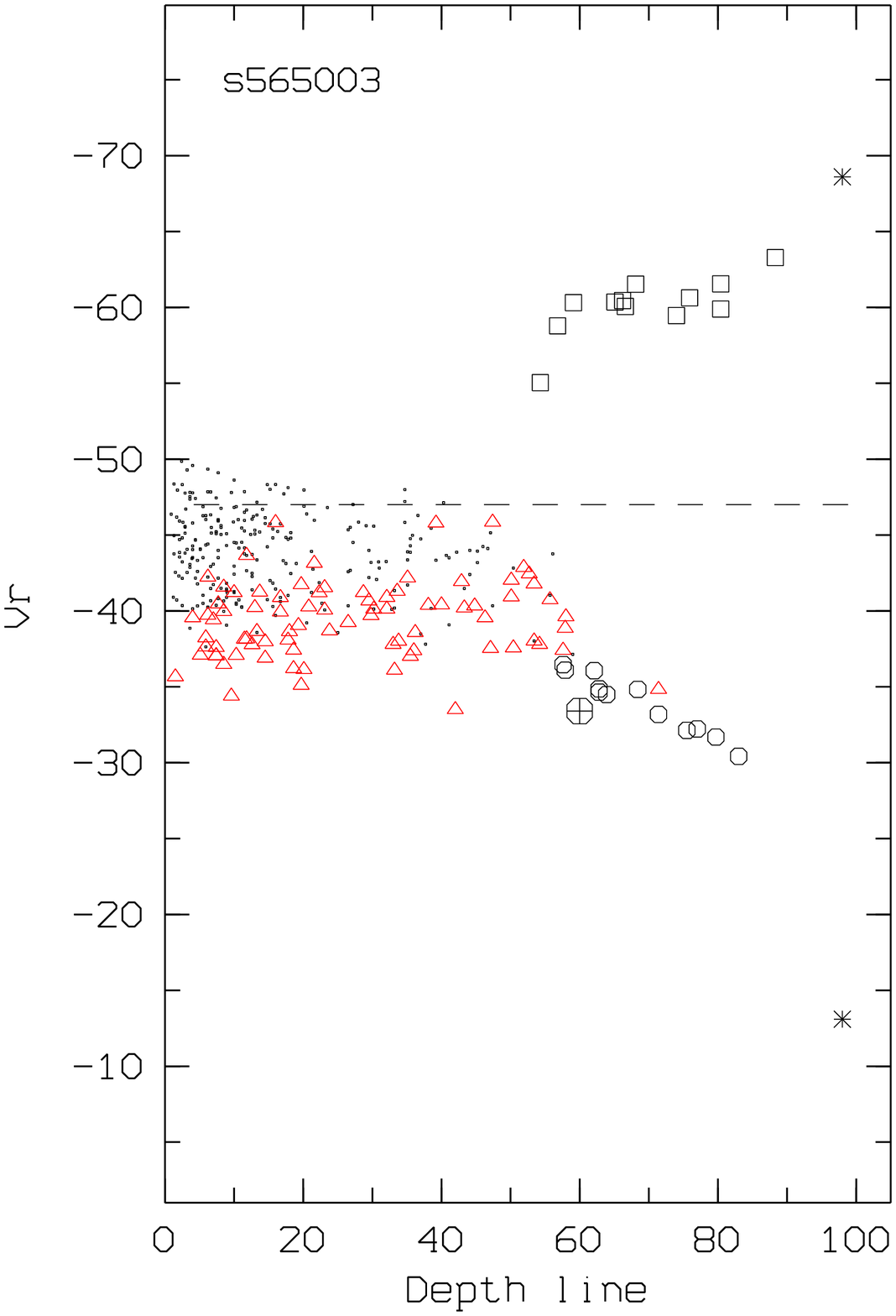} 
\includegraphics[angle=0,width=0.24\textwidth,bb=70 40 550 780,clip]{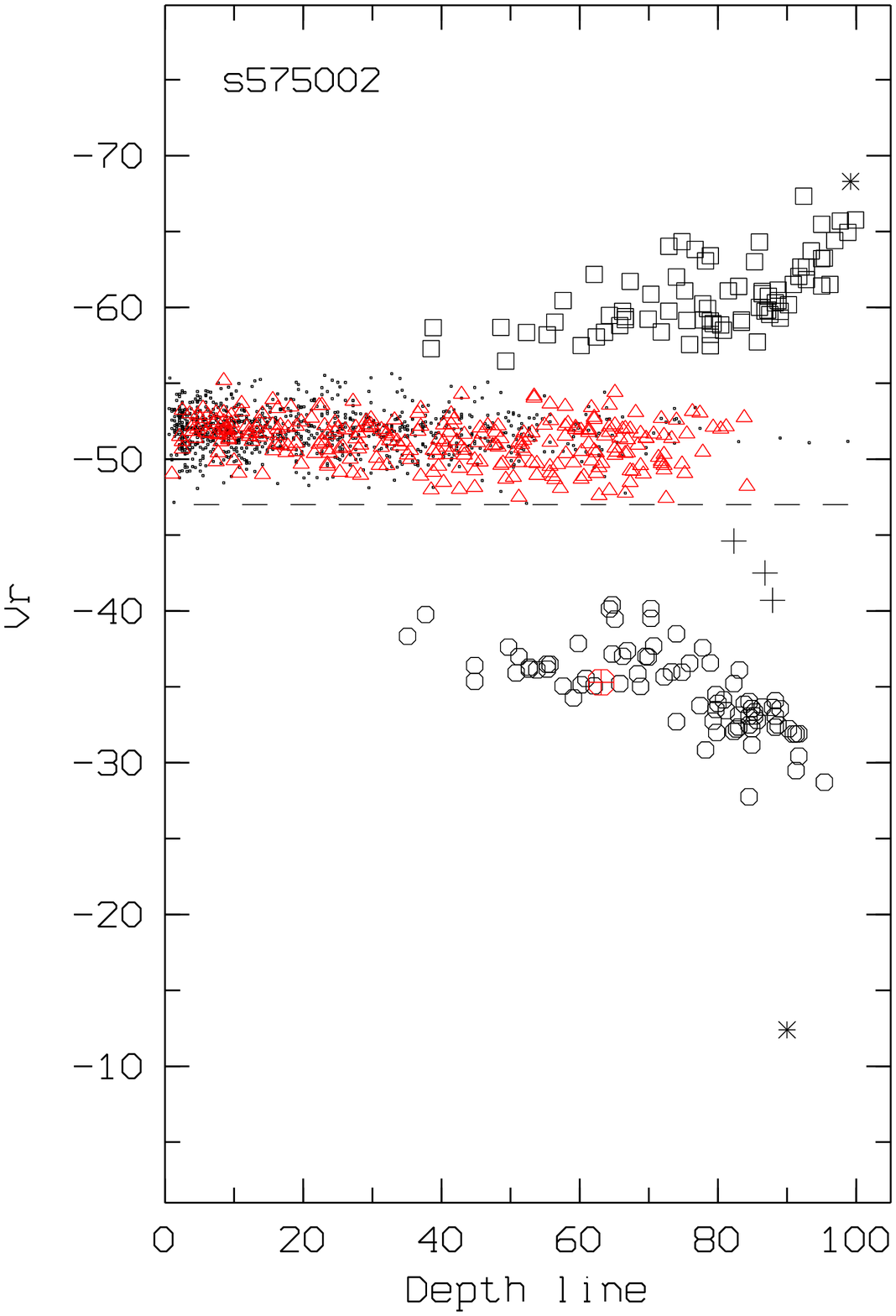} 
\caption{The radial velocity Vr measured from the absorption cores as a function of the line depth. 
        The points are data for isolated symmetric absorption lines, the triangles data for absorption lines of ions, 
        the squares and circles data for the short-wavelength and long-wavelength components of split absorption lines, 
        the crosses data for absorption components of HI lines (the H$\alpha$ line can be identified as the circled cross), 
        and the asterisks data for short-wavelength and long-wavelength components of the D\,NaI lines. 
        The dashed line indicates the systemic velocity, Vsys\,=\,$-47$\,km/s [\cite{Lobel1998,Lambert}]. 
        The arrows in the panel for the spectrum s538009 mark the velocity from the     
        FeII~6369.46 and 6432.68\,\AA{} emission components.} 
\label{fig2}  
\end{figure}

The fourth column of Table\,\ref{velocity} contains the velocity Vr(blue) averaged for lines of a given group for
each date. Three important points should be noted. First, the widths of the short-wavelength components are smaller 
than those of the long-wavelength components, and their unblended left wings have steeper gradients than do the 
right wings of the long-wavelength components. This is clearly visible in the BaII~6141\,\AA{} profile  shown in the 
upper panel of Fig.\,\ref{fig1}. Second, the positions of the short-wavelength components are close to the CO line 
core in the near-IR [\cite{Lambert,Gorlova}]. Obviously, the CO lines of an F--star can be formed only in the 
circumstellar envelope. Third, Fig.\,\ref{fig2} shows that the velocity of the short-wavelength components Vr(blue) 
is not constant. It follows from Table\,\ref{velocity} and Fig.\,\ref{fig2} that the velocity of the envelope layers 
where the envelope absorption lines are formed varies in time in a narrow range, from $-59.6$ to $-67.3$\,km/s. 
The temporal variations of the short-wavelength wings of these components (see the profiles of the BaII~6141\,\AA{} 
line in Fig.\,\ref{fig1}   and of the NaI~5889\,\AA{} line in Fig.\,\ref{fig5}) 
could be due to changes  in the wind parameters.

\begin{longtable}{l c| c}
\caption{List of split lines in the spectra of $\rho$\,Cas} 
\\ \hline 
\endfirsthead
\hline 
\multicolumn{3}{l}{Tabl.\,2, continuation} \\ \hline
\endhead
\hline
\endfoot
\hline
Element &$\lambda$, \AA{} &$\chi_{low}$, eV  \\
\hline
TiII& 3913.47 & 0.95\\ 
FeI & 3920.26 &	0.97\\ 
FeI & 3922.91 &	0.97\\ 
YII & 3950.36 &	0.90\\ 
FeI & 4005.24 &	1.56\\ 
TiII& 4012.38 &	0.57\\ 
MnI & 4030.75 &	0.00\\ 
MnI & 4033.06 &	0.00\\ 
MnI & 4034.48 &	0.00\\ 
FeI & 4045.81 &	1.48\\ 
FeI & 4063.59 &	1.56\\ 
FeI & 4071.74 &	1.61\\ 
SrII& 4077.71 &	0.00\\ 
FeI & 4132.06 &	1.61\\ 
FeI & 4143.87 &	1.56\\ 
FeI & 4202.03 &	1.48\\ 
SrII& 4215.52 &	0.00\\ 
CaI & 4226.73 &	0.00\\ 
ScII& 4246.82 &	0.32\\         
FeI & 4250.79 &	1.56\\ 
FeI & 4271.76 &	1.48\\ 
FeI & 4294.12 &	1.48\\ 
TiII& 4300.04 &	1.18\\ 
FeI & 4307.90 &	1.56\\ 
FeI & 4325.76 &	1.61\\ 
TiII& 4330.70 &	1.18\\ 
TiII& 4337.91 &	1.08\\ 
FeI & 4383.54 &	1.48\\ 
TiII& 4395.03 &	1.08\\ 
FeI & 4404.75 &	1.56\\ 
ScII& 4415.55 &	0.60\\ 
TiII& 4417.71 &	1.16\\ 
FeI & 4427.31 &	0.05\\ 
TiII& 4443.80 &	1.08\\ 
FeI & 4461.65 &	0.08\\ 
TiII& 4468.51 &	1.13\\ 
TiII& 4501.27 &	1.12\\        
TiII& 4533.96 &	1.24\\ 
TiII& 4549.62 &	1.58\\ 
BaII& 4554.03 &	0.00\\ 
TiII& 4563.76 & 1.22 \\
TiII& 4571.97 & 1.57 \\
BaII& 4934.08 & 0.00 \\
TiII& 5129.16 & 1.89 \\
FeII& 5169.03 & 2.89 \\
MgI & 5172.70 & 2.71 \\
MgI & 5183.62 & 2.71 \\
FeI & 5269.54 & 0.85 \\
FeI & 5328.04 & 0.92 \\
FeI & 5371.49 & 0.96 \\
FeI & 5397.13 & 0.92 \\
FeI & 5405.77 & 0.99 \\
FeI & 5429.70 & 0.96 \\
FeI & 5434.52 & 1.01 \\
FeI & 5446.92 & 0.99 \\
FeI & 5455.62 & 1.01 \\
BaII& 5853.67 & 0.60 \\
BaII& 6141.71 & 0.70 \\
BaII& 6496.90 & 0.60 \\   
\hline
\end{longtable}

Vr values measured from emission and absorption lines in the spectrum of $\rho$\,Cas exhibit a large
scatter, from $-10$ to $-70$\,km/s. However, it follows from Fig.\,\ref{fig2}, which displays the 
dependence of the heliocentric radial velocity Vr on the depth of the corresponding line, that the velocities measured for
each of the dates contain several fairly well isolated groups of lines. The radial velocities measured for
$\rho$\,Cas from these line groups and for our observing epochs are presented in Table\,\ref{velocity}. The numbers
of lines used to determine the mean velocity Vr for each of the groups are given in brackets. The first
group includes the vast majority of all spectral lines: isolated symmetric lines of metals that are weak or
have moderate intensity. The mean values corresponding to the positions of such absorption lines,
Vr(sym), is given in the third column of Table\,\ref{velocity}. It follows from this table and Fig.\,\ref{fig2} 
that the velocity Vr(sym), reliably measured from a large number of symmetric absorption lines, does not coincide with
the systemic velocity, and varies from epoch to epoch in the range from $-39.6$ to $-55.4$\,km/s. The mean
value, Vr(sym)\,=\,$-46.7$\,km/s, coincides with Vsys. These variations reflect pulsation motions in layers
close to the photosphere, with an amplitude of several km/s around the systemic velocity.

Our position measurements for symmetric absorption lines also show a relation between Vr and line depth on individual 
observing dates, indicating the presence of a velocity gradient in the atmosphere of the star (see, for example, data 
for the spectra taken on September 30, 2009 and September 24, 2010). In addition, a difference by 3--4\,km/s for 
velocities measured from absorption lines of neutral atoms and ions was found for some dates. This is observed for
the spectra taken on February~21, 2008 and in 2010 as well as for both spectra of January, 2011. Thus,
besides the earlier known pulsational variations, we found a weak stratification of the velocity field in the
atmosphere of $\rho$\,Cas. The difference between the velocities measured from absorption lines of neutral
atoms and ions reﬂects inhomogeneities in the outer atmospheric layers of the star. The detected differential 
line shifts vary in time, hindering a simple averaging of the velocities from arbitrary groups of
absorption lines. It is obviously necessary to study the velocity field using homogeneous groups of spectral
features, preferably those having similar intensities and ionization stages.

In addition, there is a slight velocity gradient at some epochs: Vr(blue) changes by 5--6\,km/s from
shallower to the deepest layers. The velocity gradient for the short-wavelength components of the split 
absorption lines increases the uncertainty of the mean Vr(blue) values (Table\,\ref{velocity}, the fourth column). 
Note that the velocity derived from the CO profile, which is formed in the uppermost layers of the envelope,
continues the trend for increasing envelope expansion velocity shown by the strong absorption lines.

\clearpage

We are the first to detect a velocity gradient in the envelope. The possible presence of a gradient in the
outﬂow velocity was noted by Sargent [\cite{Sargent}]. Later, Lambert et al. [\cite{Lambert}] used the possible 
presence of a velocity gradient in the stellar envelope to explain the asymmetry of the CO profile.

The third group of lines is comprised of the long-wavelength components of split absorption lines.
Figure\,\ref{fig2} shows that, for most of these spectra, the dependence of Vr on the line depth is a continuation
of the relations plotted for symmetric absorption lines. Note that the uncertainties of the mean velocities are
much higher for the long-wavelength (Table\,\ref{velocity}, fifth column) than for the short-wavelength components.
This is probably due to the complex character of the profiles of the long-wavelength components. We
consider this in more detail below.

\begin{table}
\caption{Heliocentric radial velocities of $\rho$\,Cas. Vr(sym) is the mean velocity from symmetric absorption lines; 
         Vr(blue) and Vr (red) are the mean velocities from the short-wavelength and long-wavelength components of split 
         absorptions. The number of lines used to determine the mean values is in brackets; the last column 
         contains the Vr(HI$_{-}$abs) velocity measured from neutral-hydrogen absorption cores}
\begin{tabular}{c|c|l|l|l|l}
\hline
Spectrum &Date & \multicolumn{4}{c}{\small Vr,km/s} \\
\cline{3-6}&  &\hspace{0.9cm}sym&\hspace{0.9cm} blue &\hspace{0.9cm} red & HI$_{-}$abs \\
\hline     
 s493015& 09.03.2007&$-55.1\pm0.09$\,(290) &$-65.5\pm 0.4$\,(12)&$-34.9\pm 0.4$\,(12)&$-46.5^1$  \\
 s494023& 10.03.2007&$-55.4\pm0.10$\,(228) &$-66.9\pm 0.4$\,(16)&$-35.0\pm 0.4$\,(16)&$-47.4^1$  \\
 s495019& 10.03.2007&$-55.0\pm0.09$\,(296) &$-67.3\pm 0.4$\,(17)&$-34.1\pm 0.4$\,(16)&$-46.3^1$	 \\
 s516015& 21.02.2008&$-44.0\pm0.11$\,(217) &$-61.7\pm 0.5$\,(12)&$-37.3\pm 0.5$\,(11)&$-40.0^2$  \\
 s525032& 19.10.2008&$-40.9\pm0.07$\,(259) &$-65.8\pm 0.1$\,(89)&$-34.4\pm 0.2$\,(83)&$-49.1^3$, $-45.5^4$\\
 s526006& 20.10.2008&$-51.1\pm0.09$\,(235) &$-62.2\pm 0.4$\,(12)&$-37.1\pm 0.5$\,(12)&$-44.6^2$   \\       
 s538009& 30.09.2009&$-43.6\pm0.11$\,(220) &$-62.8\pm 0.3$\,(12)&$-34.3\pm 0.4$\,(12)&$-46.2^2$   \\
 s553018& 01.08.2010&$-43.4\pm0.11$\,(170) &$-61.8\pm 0.3$\,(11)&$-32.6\pm 0.3$\,(11)&$-40.2^1$	  \\ 
 s554032& 23.09.2010&$-39.6\pm0.08$\,(333) &$-60.0\pm 0.2$\,(41)&$-30.0\pm 0.3$\,(34)&$-36.4^1$, $-36.4^3$, $-35.0^4$ \\  
 s555027& 24.09.2010&$-40.2\pm0.12$\,(173) &$-60.3\pm 0.4$\,(11)&$-33.4\pm0.4$\,(12) &$-30.2^2$   \\
 s564020& 13.01.2011&$-44.0\pm0.10$\,(235) &$-59.6\pm 0.3$\,(11)&$-32.9\pm0.4$\,(12) &$-32.6^2$   \\
 s565003& 13.01.2011&$-43.9\pm0.10$\,(249) &$-60.1\pm 0.4$\,(12)&$-33.9\pm0.4$\,(12) &$-33.4^2$   \\       
 s575002& 14.09.2011&$-51.8\pm0.04$\,(934) &$-60.7\pm 0.2$\,(76)&$-34.8\pm0.2$\,(82) &$-52.9^1$, $-29.8^2$, $-54.4^3$, \\
        &           &                      &                    &                    &$-49.4^4$  \\
\hline
\multicolumn{6}{l}{\footnotesize {\it Vr from neutral-hydrogen absorption cores: $^1$ -- H$\beta$, $^2$ -- H$\alpha$,
                  $^3$ --  H$\delta$, $^4$ --  H$\gamma$.}}
\end{tabular}   
\label{velocity} 
\end{table}

\begin{figure}
\includegraphics[angle=0,width=0.54\textwidth,bb=10 40 550 780,clip]{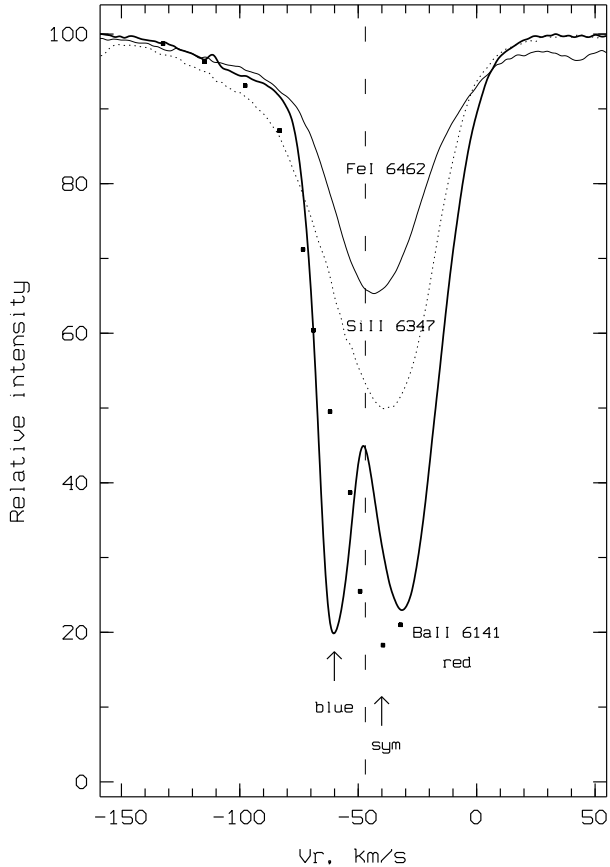} 
\caption{FeI~6462\,\AA{}, line profiles in the spectrum of $\rho$\,Cas obtained on September~24, 2010. 
    The vertical arrows mark the mean (for this date) velocities derived from symmetric absorption lines, 
    Vr(sym)$\approx -40.1$\,km/s, and from short-wavelength components, Vr(blue)\,$\approx -60.3$\,km/s.
    The vertical dashed line is the systemic velocity, Vsys\,=$-47$km/s [\cite{Lambert,Lobel1998}]. 
    The points show the assumed shape  of the wing of the BaII~6141\,\AA{} photospheric 
    line in the absence of emission. The position of this line’s core in the diagram corresponds to the positions
    of symmetric moderate-intensity absorption lines in the same spectrum.} 
\label{fig3}  
\end{figure}

Lambert et al. [\cite{Lambert}] noted that the profiles of resonance (and strongest subordinate) 
photospheric absorption lines in the spectrum of $\rho$\,Cas were complex as early as 1981. 
We also believe that the long-wavelength components of the split absorption lines are ordinary strong 
absorption lines formed in high atmospheric layers. Their radial velocities differ from those of the 
high-intensity, isolated absorption lines due to the velocity gradient, as well as those of
stationary emission lines formed in the circumstellar medium. The circumstellar emission lines, 
whose positions in the spectrum of $\rho$\,Cas do not coincide with the systemic velocity, exceed the local continuum
level during outburst [\cite{Lobel}]. For observations outside outburst, the emission lines are weak, 
but they distort the long-wavelength components, shifting their cores toward longer wavelength. Figure\,\ref{fig3} 
presents the profiles of several absorption lines with different intensities in the spectrum obtained on 
September~24, 2010 by way of illustration. The positions of absorption cores with low (FeI~6462\,\AA{}) and moderate
intensity (SiII\,6347\,\AA{}) correspond to the mean velocity for this epoch, Vr(sym)\,=$-40$\,km/s 
from Table\,\ref{velocity}.
At the same time, the core of the strong BaII\,6141\,\AA{} line is shifted by approximately +8\,km/s with respect
to the relatively weak absorption lines.

\begin{figure}[t]
\includegraphics[angle=-90,width=0.8\textwidth,bb=10 40 550 780,clip]{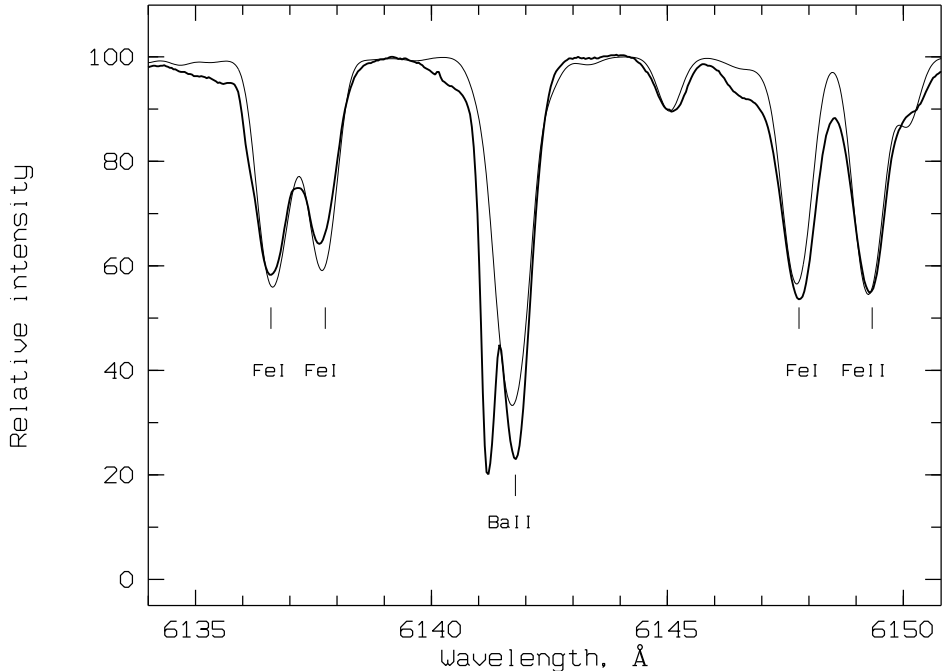}  
\caption{Fragment of the spectrum of $\rho$\,Cas obtained on September~24, 2010 (JD 2455654.4; bold curve) 
        compared to the theoretical spectrum computed for Teff\,=\,6000\,K, $\log g$\,=\,0.1, microturbulence 
        velocity $\xi_t$\,=\,12\,km/s, and macroturbulence velocity $\zeta_t$\,=\,20\,km/s (thin curve). }
\label{fig4}  
\end{figure}

Figure\,\ref{fig4} compares a fragment of the observed (September~24,~2010) and synthetic spectra containing 
absorption lines with various intensities, including the split BaII\,6141\,\AA{}  line.  The calculations were
done using the modified STARSP code [\cite{Tsymbal}], adapted for a Linux environment. According to the data from
Table\,\ref{obs}, the effective temperature for September~24, 2010 was Teff$\approx 6000$\,K. We took the surface gravity
$\log g = 0.1$, microturbulence velocity $\xi_t$\,=\,12\,km/s, and slight deviations of the chemical abundances
from the solar values from [\cite{Takeda1998}]. Theoretical modeling of spectra does not provide the necessary accuracy
for extremely luminous stars. However, we did not attempt to accurately reproduce the line intensities,
since we were mainly interested in their positions.
The position of the theoretical BaII\,6141\,\AA{} line profile agrees well with the long-wavelength component
in the observed spectrum. This provides additional evidence supporting our suggestion that the long-wavelength 
components of split absorption lines are formed in the same layers of the stellar atmosphere as
those where the symmetric, isolated absorption lines are formed. The short-wavelength component clearly
originates in the envelope.

Note that splitting of strong absorption lines was detected [\cite{Klochkova2009}] in the spectrum of 
the post-AGB supergiant V354\,Lac, whose mass and evolutionary stage are very different from those of 
the hypergiant $\rho$\,Cas. Analysis of radial-velocity data from echelle spectra of V354\,Lac obtained using 
the 6-m telescope\,+\,the NES spectrograph shows that the strongest absorption lines of metal ions 
(BaII, LaII, CeII, NdII) are distorted by a component formed in an envelope expanding at a constant velocity.

Gesicki [\cite{Gesicki}] used photographic spectra of $\rho$\,Cas taken by J.~Smolinski during 1969--1970 to study
the temporal behavior of the BaII\,4934, 5853 and FeI\,5328\,\AA{} lines. The main conclusion of this work
was that the short-wavelength component is formed in an expanding circumstellar envelope with a very
high gas temperature, $\approx$12000\,K. Our results are consistent with this formation region for the 
short-wavelength component. However, Gesicki [\cite{Gesicki}] concluded from monitoring of a small number of lines that
neither component of the split absorption lines varied their velocity Vr during the pulsation period. Our
more accurate measurements demonstrate that the positions of both components of the split absorption
lines vary in time with an amplitude of several km/s.

Recently, Gorlova et al. [\cite{Gorlova}] suggested a different explanation of the splitting of low-excitation 
absorption lines observed in spectra of $\rho$\,Cas. They suggest that the splitting is due to a stationary 
emission feature with Vr$\approx -50$\,km/s formed in the circumstellar medium, which overlaps with absorption 
lines broadened by strong turbulence. Their main argument against the hypothesis that the short-wavelength
components are formed in the circumstellar medium is that the presence of two independent absorption
components would require the continual existence of layers moving both outwards and inwards towards
the stellar center, which can be ruled out for physical reasons.

\begin{figure}[hbtp]
\includegraphics[angle=0,width=0.6\textwidth,bb=10 40 570 790,clip]{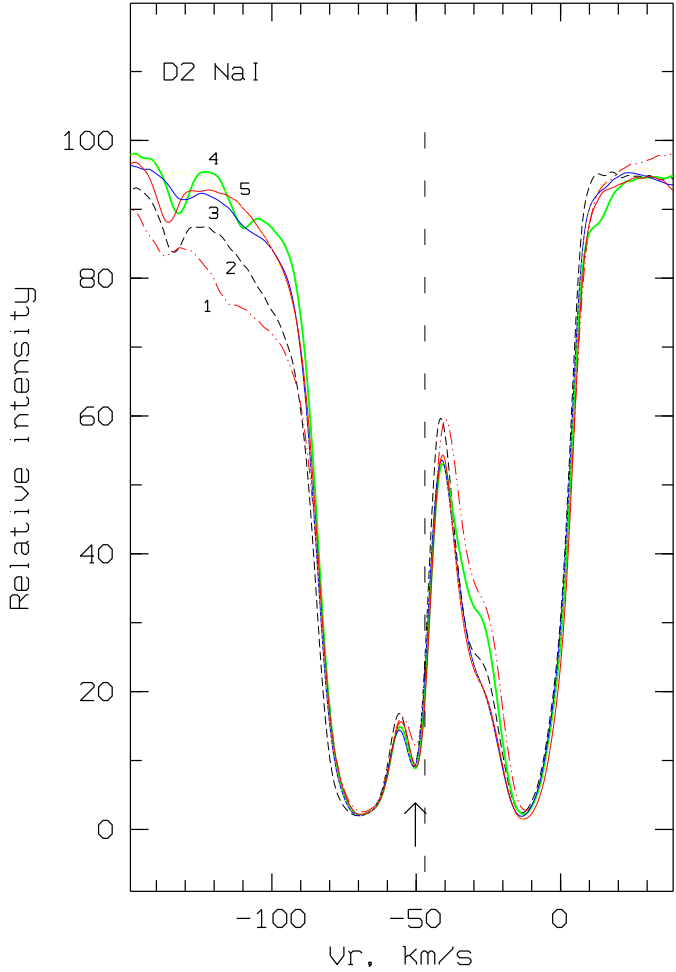} 
\caption{Profile of the D2~NaI line in the spectra of $\rho$\,Cas at various epochs: 1 -- February~21, 2008; 
         2 -- October~20, 2008;  3 -- September~30, 2009; 4 -- September~24, 2010; 5 -- January~13, 2011. 
         The arrow marks the interstellar component, V(IS)$\approx -50$\,km/s [\cite{Georgelin}]. 
         The vertical dashed line is the systemic velocity,  Vsys\,=\,$-47$\,km/c~[\cite{Lambert,Lobel1998}].} 
\label{fig5}  
\end{figure}

Note that the assumption of oppositely directed motions (compression of some layers and expansion
of others) is justified in the 1D model atmosphere if both the zones of compression and expansion have
densities and temperatures sufficient to form absorption lines with detectable intensities. In this case,
lines due to different ionization states or excitation states can dominate in different zones. 
If the oppositely directed motions are present in different parts of the observed stellar hemisphere 
(for example, due to non-symmetric pulsations), the condition for the line components formed in the 
ascending and descending parts of the hemisphere to be simultaneously detectable is determined by the ratio 
of the combined areas of the ascending and descending parts and the ratio of the corresponding temperatures 
(since the contribution to the flux is dependent on the temperature as well as the area).

Our proposed explanation for the observed splitting of the strongest absorption line suggests a
simpler pattern for the differential motions: low-amplitude pulsations plus expansion of the upper
layers of the extended atmosphere forming a transition to the envelope, plus the presence of velocity
gradients in the stellar atmosphere and envelope at some times. Another important difference of our
results from those of Lobel et al. [\cite{Lobel1994,Lobel1998}] or Gorlova et~al. [\cite{Gorlova}] 
is that, using spectra over a wide wavelength range, we have been able to analyze velocities from 
symmetric absorption lines whose profiles are not distorted by circumstellar features. As a result, the
pulsation amplitude we have found, about $\pm7$\,km/s, is half the value derived in [\cite{Lobel1994}].

\section*{Other lines with anomalies.}

In addition to broad components similar to those we see for the strongest absorption lines, the D~NaI line
profiles (Fig.\,\ref{fig5}) contain a narrow absorption feature at Vr\,=\,$-49.8$\,km/s and a component near 
Vr$\approx -29$\,km/s, poorly visible in our spectra. The absorption at Vr$\approx -50$\,km/s has an 
interstellar origin, and corresponds to the location of $\rho$\,Cas in the Perseus arm [\cite{Georgelin}]. 
The interstellar NaI line corresponding to the Local Arm cannot be identified against the long-wavelength 
photospheric component.

In the spectra s516015 and s538009, we measured an abnormal Vr velocity from the FeII~6369.46 and 6432.68\,\AA{}
ion lines. When these spectra were obtained, the short-wavelength wings of the two lines,
which have the same ionization potential $\chi_{low}$\,=\,2.89\,eV, were distorted by emission, especially clearly
expressed in the s538009 spectrum. The horizontal arrow in the corresponding panel of Fig.\,\ref{fig2} 
indicates the mean velocity of the emission component of the FeII~6369.46 and 6432.68\,\AA{} lines in these spectra,
Vr\,=$-68.4$\,km/s. Obviously, these wind emission lines were formed in the same layers as the short-wavelength 
components of the split metallic absorptions.

In addition to the position measurements for the groups of metal lines noted above, we also present the
velocity derived from the absorption components of the H$\delta$, H$\gamma$, H$\beta$, and H$\alpha$ 
lines for all the spectra containing neutral-hydrogen lines. The positions of the H$\delta$, H$\gamma$, and 
H$\beta$ cores, plotted as crosses in Fig.\,\ref{fig2}, vary between the systemic velocity and the
velocity derived from the absorption lines of metal ions (triangles). At some epochs, the position of the 
H$\alpha$ absorption core (circled cross) also coincides with the systemic velocity, but, more often, this 
line follows the Vr(r) sequence for the long-wavelength components of split absorption lines.

\section{Conclusions}

We have derived effective temperatures and radial velocities from spectral features formed in different 
layers of the extended atmosphere of the hypergiant $\rho$\,Cas using 12 high-quality echelle spectra 
taken during various observing seasons in 2007--2011. The effective temperature  of the star varied within
5777--6744\,K.

Due to our wide wavelength range, we were able to study the velocity field using an unprecedentedly large
number of isolated (several hundred in each spectra) and split (from 12 in the visual to 89 in 
the short-wavelength range) absorptions. The radial velocity derived from weak, symmetric absorption lines of
metals varies from epoch to epoch with an amplitude of about $\pm7$\,km/s relative to Vsys\,=\,$-47$\,km/s, as a
consequence of low-amplitude pulsations in the stellar atmosphere. At certain times, we observed a relation
between the radial velocity and the line intensity, indicating the presence of a velocity gradient in deep
layers of the stellar atmosphere. For several phases, we also found a difference of 3--4\,km/s between the
velocities measured from absorption lines of neutral atoms and of ions. Thus, we are the first to detect
velocity stratification in the atmosphere of $\rho$\,Cas.

We demonstrated that the long-wavelength components of the split BaII, SrII, TiII absorptions
and of other strong lines with low excitation potentials for their lower levels were distorted by a stationary
emission feature, which shifted the line toward longer wavelengths. Thus, the long-wavelength components of 
the split absorption lines are ordinary absorption features; taking into account the distortion by the 
stationary emission feature at individual epochs, the formation region and radial velocities of these 
components do not differ from those for single absorption lines.

The radial velocities of short-wavelength components were reliably determined using a large number
of features, and lie in a narrow range from Vr(blue)$\approx -60$ to $-70$\,km/s. The short-wavelength 
components of these absorption lines are formed in the circumstellar envelope, where one component of the
D\,NaI doublet lines and the emission components  for the FeII\,6369.46 and 6432.68\,\AA{} ion lines are 
also formed.

\section*{Acknowledgments}

The authors thank Dr.~M.V.~Yushkin for his great help during the observations.  
This study was supported  by the Russian Foundation for Basic Research (project no.~11--02--00319\,a).

\end{document}